\documentclass{appolb}
\usepackage{epsfig}
\newcommand{\pt}{$p_T$ }

\begin{document}
\title{Identified Particle Correlations at RHIC: Medium Interactions \&
Modified Fragmentation
}
\author{Anne Sickles
\address{Brookhaven National Laboratory}
}
\maketitle
\begin{abstract}
Azimuthal angle two particle correlations have been shown to be a 
powerful probe for extracting novel features of jet induced correlations
produced in Au+Au collisions at RHIC.  At intermediate $p_T$, 2-5GeV/c,
the jets have been shown to be significantly modified in both their
particle composition and their angular distribution compared to p+p
collisions.  Two-particle angular correlations with identified particles 
provide sensitive probes of both the interactions between hard scattered 
partons and the medium. 
The systematics of these correlations are essential 
to understanding the physics of intermediate $p_T$ in heavy ion collisions.
\end{abstract}
  
In $p+p$ collisions at $\sqrt{s}$=200GeV a hard scattering picture has
been found to well describe single particle spectra of both pions~\cite{ppg024,starxt}
and $p$ and $\bar{p}$ at \pt as low as 2GeV/$c$.  In heavy ion collisions
at the same energy, it is known from direct photon data~\cite{isobeqm} that the initial hard parton
parton collisions scale as expected 
from $p+p$ by the number of binary nucleon-nucleon
collisions.
There is also evidence that the soft physics region
extends higher in $p_T$.  The baryon to meson ratios are increased by $\approx$3 in
central Au+Au collisions compared to $p+p$ collisions for 2$<p_T<$5GeV/$c$~\cite{ppg030} 
and a quark number scaling of $v_2$ has been found to extend to $\approx$4-6GeV/$c$~\cite{ppg062}.

Because of this overlap,
intermediate \pt is well suited to studying the interaction between the calibrated hard
probes and the strongly interacting medium.
Jet-induced two particle correlations have been used extensively to probe this physics,
and have led to many unexpected results at RHIC such as jet-like structure
of the intermediate \pt baryon excess~\cite{ppg033} and the away side jet shape has a peak
displaced from 
azimuthal angular difference, $\Delta\phi$,  of 
$\pi$~\cite{ppg067}.  We present results which exploit the high 
statistics and particle identification capabilities of RHIC detectors.

\begin{figure}[t]
\begin{minipage}{0.5\textwidth}
\includegraphics[width=\textwidth]{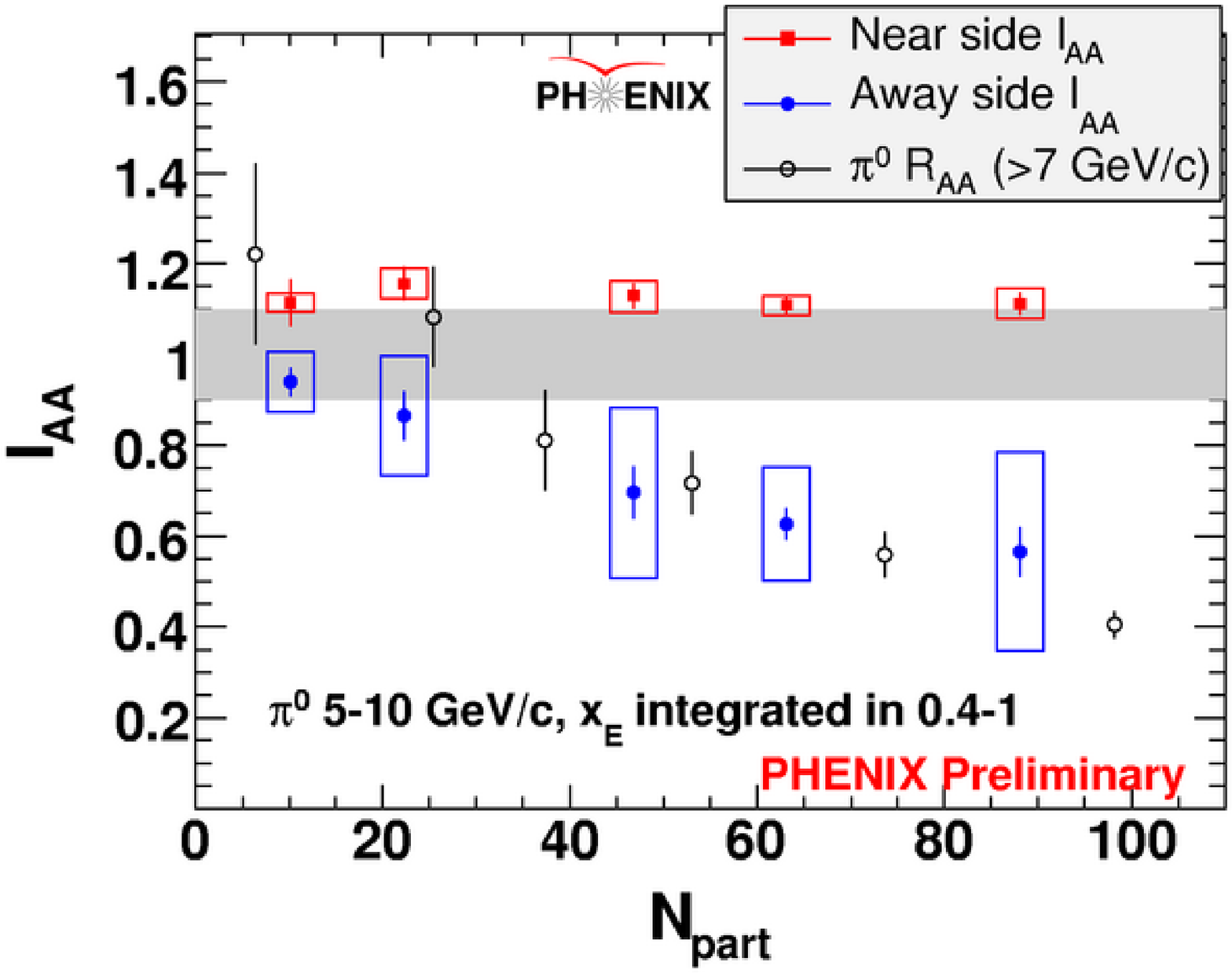}
\end{minipage}
\hspace{\fill}
\begin{minipage}{0.5\textwidth}
\includegraphics[width=\textwidth]{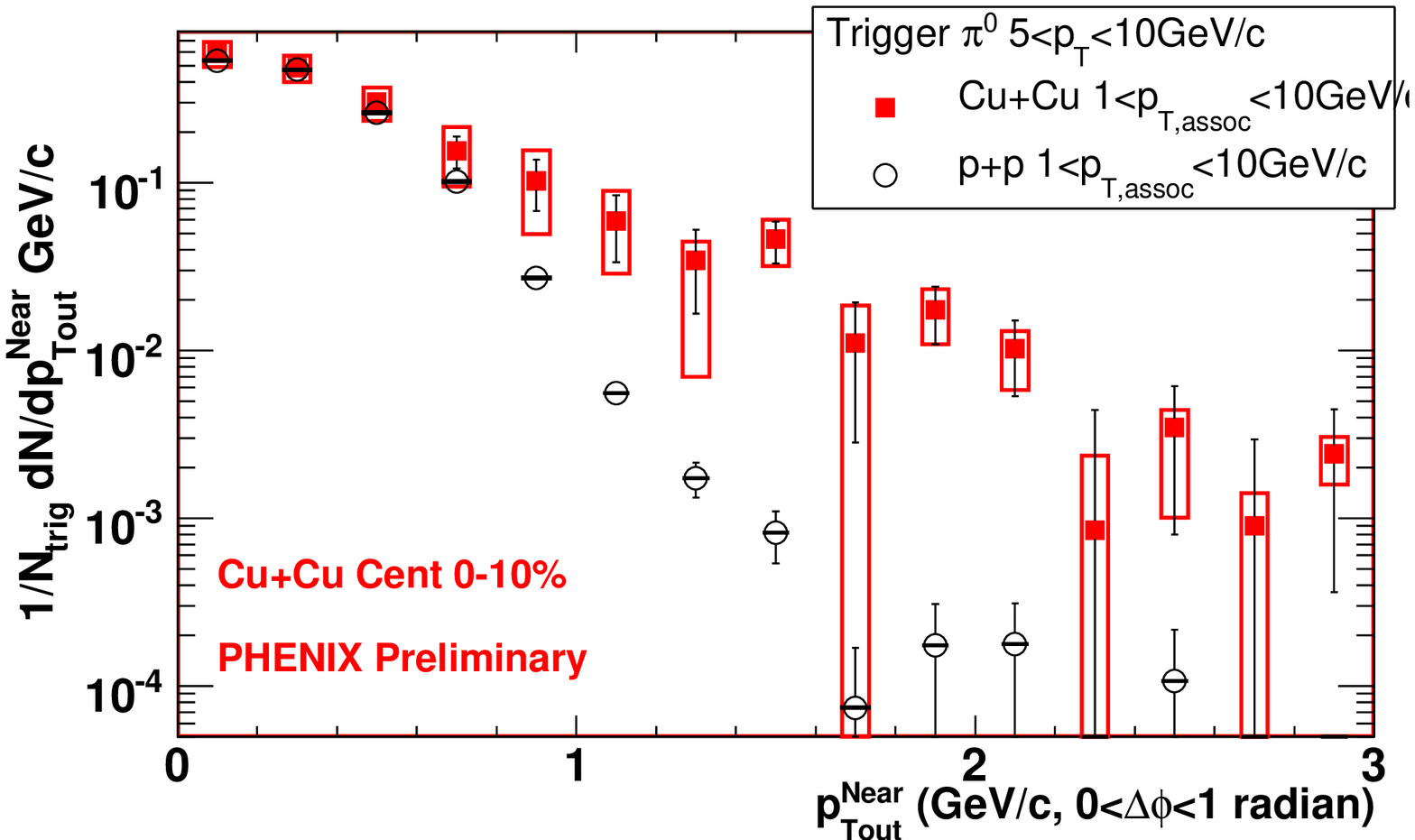}
\end{minipage}
\caption{(left) $I_{AA}$ as a function of $N_{part}$ for $\pi^0$-hadron correlations in Cu+Cu collisions.
(right) $p_{out}$ distribution for central Cu+Cu and p+p $\pi^0$-hadron correlations.  In both panels
trigger $\pi^0$ are 5$<p_T<$10GeV/$c$.}
\label{figcucu}
\end{figure}

Due to the large energy loss, at high \pt the hard scattered partons are expected to emerge 
from near the
surface of the interaction region and fragment in vacuum~\cite{wangplb}. 
Fig. \ref{figcucu} (left)~\cite{jiaqm}
shows near side, small $\Delta\phi$, $I_{AA}$  for
 trigger $\pi^0$ and lower \pt associated hadrons 
as a function of the number of participating nucleons, $N_{part}$.
$I_{AA}$ is the yield of associated hadrons per
trigger in Cu+Cu collisions divided by that in p+p collisions.  In the absence of 
nuclear effects $I_{AA}$ will be 1.  
The near side $I_{AA}$ is consistent with 1 within systematics at all centralities.
A more differential look at the fragmentation differences between Cu+Cu and p+p can be
obtained by looking at the yield as a function of $p_{out}=p_{T,assoc}\sin (\Delta\phi)$,
Fig.~\ref{figcucu} (right)~\cite{peisqm}.
Though $I_{AA}$, dominated by small $p_{out}$,
 shows no significant nuclear effects, the $p_{out}$ distributions are much
harder for Cu+Cu collisions. 
These two plots show the need for sensitive observables
to measure subtle changes to jet fragmentation.  

\begin{figure}[t]
\begin{minipage}{0.5\textwidth}
\includegraphics[width=\textwidth]{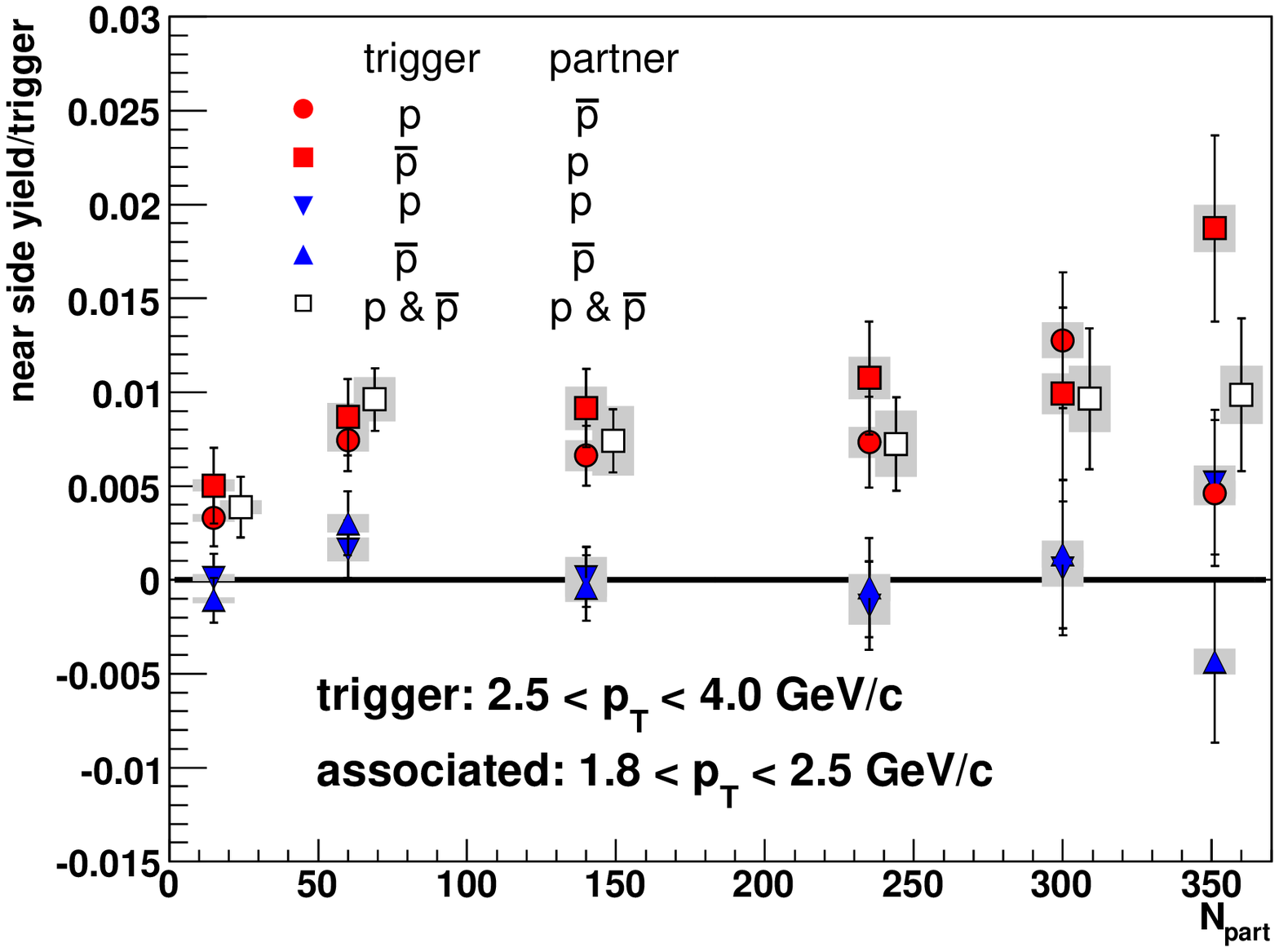}
\end{minipage}
\hspace{\fill}
\begin{minipage}{0.5\textwidth}
\includegraphics[width=\textwidth]{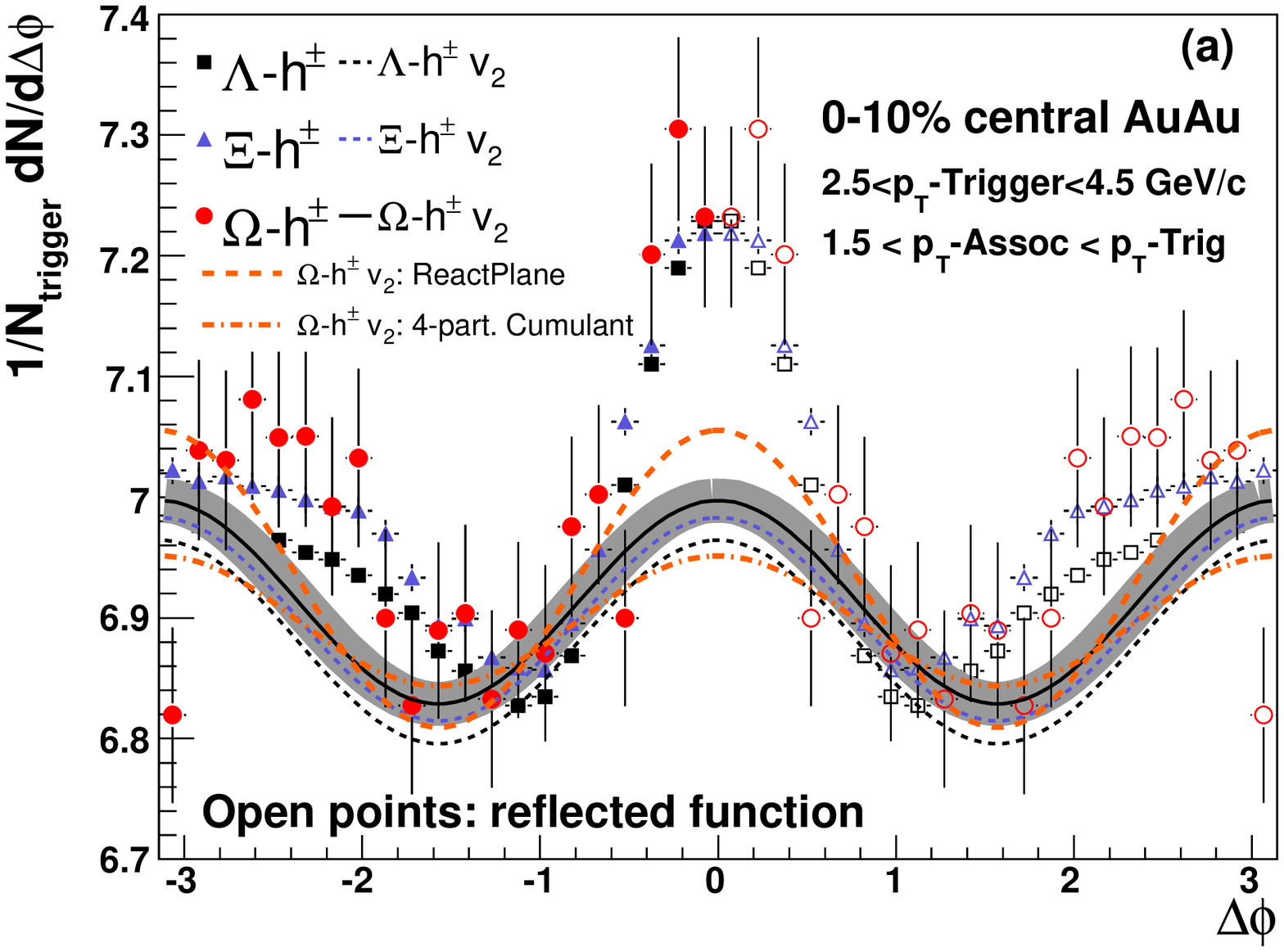}
\end{minipage}
\caption{(left) Near side correlations between two $p$ and $\bar{p}$ as a function
of $N_{part}$.  (right) Azimuthal correlations between strange hadron triggers
and hadrons.  No $v_2$ subtraction has been done.}
\label{figid}
\end{figure}

At lower \pt where there are known modifications to particle production via vacuum 
fragmentation~\cite{ppg033,starlowpt,ppg032} conditional 
yield measurements as a function of $N_{part}$ can be used
along with single particle spectra measurements constrain the modifications. 
Fig. \ref{figid} (left) shows correlations between two  $p$ or
$\bar{p}$~\cite{ppg072}.  
In the case of minimal nuclear effects, $N_{part}$=15,
the conditional yield for opposite sign pairs is independent of whether the trigger is a 
$p$ or $\bar{p}$ and the conditional yield for same sign pairs is consistent with zero.
Both features remain at all centralities, however, the
 yield of opposite sign pairs increases slightly from  
 peripheral collisions.    

It was thought the expected high 
rate of gluon conversion to $s\bar{s}$ 
in heavy ion collisions would enhance soft $s\bar{s}$ pairs  
suppressing jet-like correlations with increasing hadronic strangeness 
content~\cite{rudystrange}.  
Fig.~\ref{figid} (right)~\cite{janaqm} shows no significant difference
between the jet-like correlations triggered by $\Lambda$, $\Xi$ or $\Omega$.
Comparisons to p+p collisions are needed to quantify
differences as a function of strangeness in a purely fragmentation
system.

\begin{figure}[t]
\begin{minipage}{0.5\textwidth}
\includegraphics[width=\textwidth]{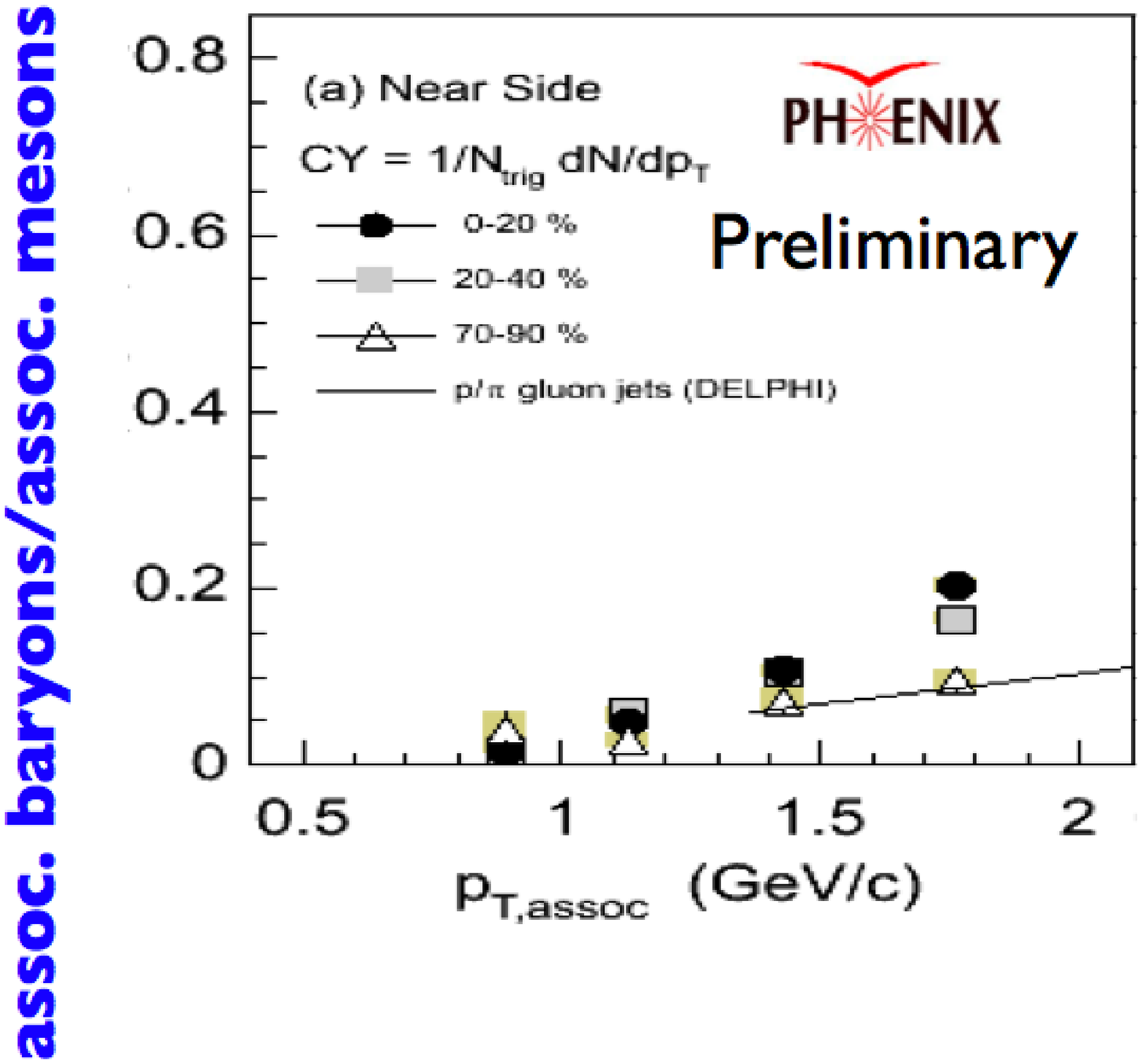}
\end{minipage}
\hspace{\fill}
\begin{minipage}{0.5\textwidth}
\includegraphics[width=\textwidth]{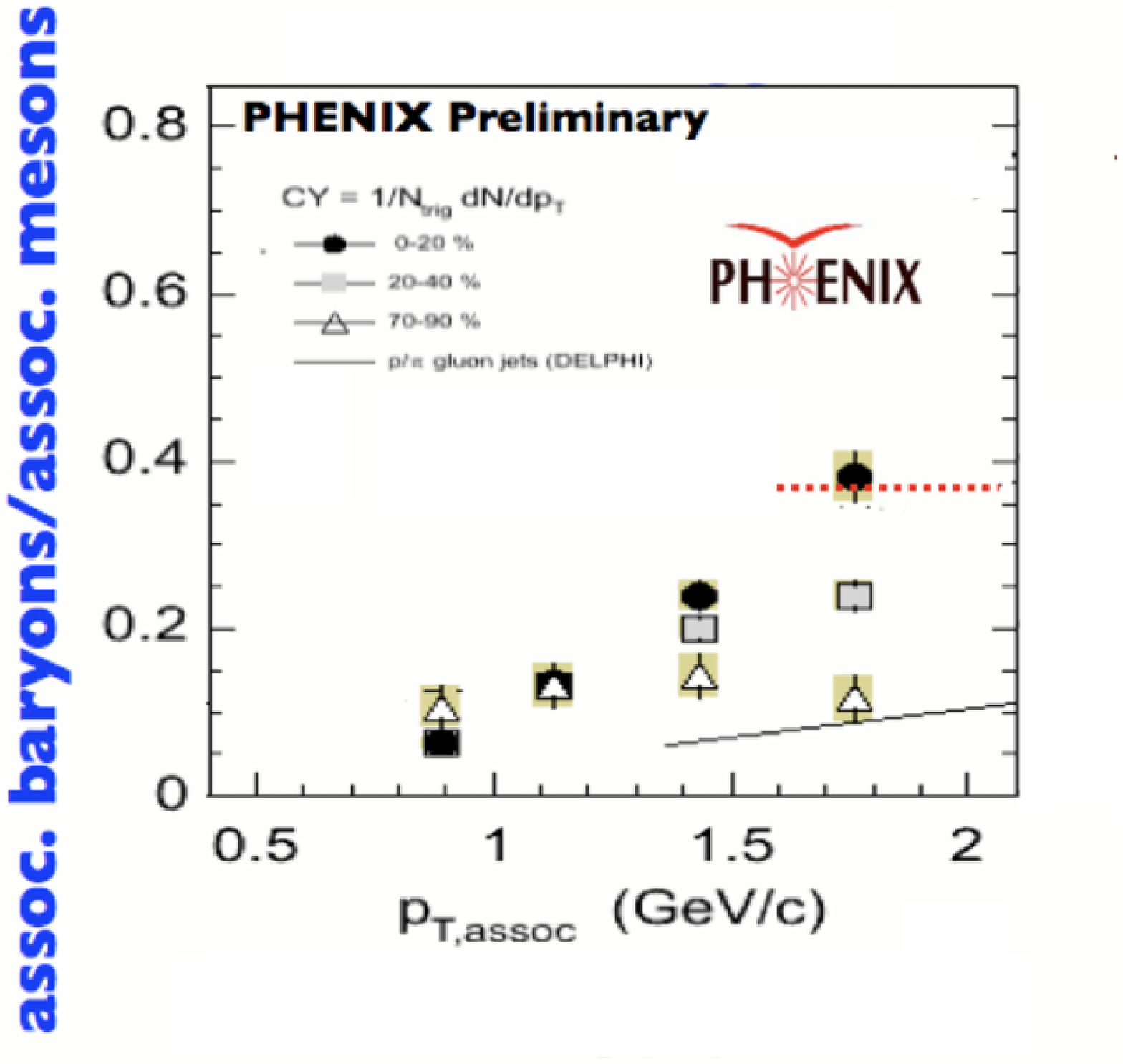}
\end{minipage}
\caption{Associated particle baryon to meson ratios as a function of $p_T$i for
near (left) and away (right) side.  Trigger
particles are hadrons, 2.5$<p_T<$4.0GeV/$c$.}
\label{fig34}
\end{figure}

Perhaps the most unexpected intermediate \pt result is the modified away side
shape observed in hadron-hadron, hadron-meson and hadron-baryon 
azimuthal correlations~\cite{ppg032,wolfhp,zuosqm}.  Models which include
a strong medium response to the recoiling parton, including Mach cones, have
been successful in describing the away side shape~\cite{ppg067}.  Because of
the near side surface bias,
the away side might be biased toward longer than average medium 
path lengths, and thus more sensitive to medium effects.  
Fig.~\ref{fig34} shows the near (left)
and away (right) side associated baryon to meson ratios as a function of the associated
particle $p_{T}$ for three centralities.  Both sides show increasing baryon to meson
ratios with centrality, however the increase is much stronger on the away side.  The
dashed line in Fig.~\ref{fig34} (right) shows the inclusive baryon to meson
ratio for 0-20\% centrality and $p_T$ of 1.85GeV/$c$.  The agreement between the 
jet associated baryon to meson ratio and the inclusive ratio could indicate that
baryons in both cases are enhanced by a common mechanism.  

We have presented some recent results of identified particle 
jet correlations from RHIC.  
At intermediate $p_T$, observations from two particle correlations
show features characteristic of both hard and soft physics.  
Correlations between high \pt $\pi^0$ and lower \pt hadrons are observed
to be modified in Cu+Cu collisions from p+p collisions.
At intermediate $p_T$, correlations between two $p$ and $\bar{p}$ retain the same charge
ordering at all centrality, indicating correlated $p$ and $\bar{p}$ production.
In central collisions the yield of hadrons associated with strange
baryons is independent of the strangeness content.
  These results suggest that, at least some of, the
baryon excess is connected to jet fragmentation in central
Au+Au collisions being modified compared to vacuum fragmentation.
These observations, along with the quark number scaling observed in elliptic
flow measurements in the same \pt range could indicate that 
particle production at intermediate \pt is a novel interplay
of hard and soft physics.  A full understanding of this
phenomenology will require models which are able to 
simultaneously explain single particle yields,
elliptic flow and jet correlations.

\bibliographystyle{iopart-num.bst}
\bibliography{sickles_ismd07}
\end{document}